\documentclass[paper,nofootinbib] {revtex4}

\usepackage{graphicx}
\usepackage{epsfig}
\usepackage{fancyhdr}
\usepackage{empheq}
\usepackage{mathbbol}
\usepackage{wasysym}
\usepackage{pstricks}
\usepackage{color}
\usepackage{bbm}
\usepackage{multirow}
\def\jp{J/\psi}
\def\ppc{\pi^+\pi^-}	

\def\etal{\textit{et al.}}

\def\be{\begin{equation}}
\def\ee{\end{equation}}

\begin{document}

\title{The momentum distribution of $J/\psi$ in $B$ decays}

%

\author{T. J. Burns$^{\P}$,  F. Piccinini$^{\dag}$, A. D. Polosa$^\ddag$, V. Prosperi$^\ddag$ and C. Sabelli$^{\ddag,\P}$}
\affiliation{
$^\P$INFN Roma, Piazzale A. Moro 2, Roma, I-00185, Italy\\
$^\dag$INFN Pavia, Via A. Bassi 6, Pavia, I-27100, Italy\\
$^\ddag$Department of Physics, Universit\`a di Roma, `Sapienza', Piazzale A. Moro 2, Roma, I-00185, Italy}

\begin{abstract}
The discrepancy between theory and data in the momentum distribution of slow $J/\psi$ in $B$ decays has been 
several times addressed as a puzzle. Using the most recent results on exclusive $B$ decays  into $\jp$ and heavy kaons  or exotic mesons  and reconsidering the non-relativistic-QCD calculation of the color octet fragmentation 
component, we show that an improvement in the comparison between data and theory can be obtained.
There is still room for a better fit to data and this may imply that new exotic mesons of the $XYZ$ kind have yet to be discovered.    
\\ \\
PACS: 14.40.Rt, 13.85.Ni, 13.25.Hw.
\end{abstract}

\maketitle

\thispagestyle{fancy}

In the first study of the inclusive decays of $B$ mesons to $\jp$, the CLEO collaboration~\cite{Balest:1994jf} observed that the  $\jp$ momentum  
distribution $d\Gamma/d{p_{\psi}}$ is well described 
at large $p_\psi$ by the sum of the exclusive modes $B\to K\jp $ and $B\to K^* \jp$~\cite{Alam:1994bi}. At lower momenta there are significant contributions from feed-down processes $B\to\chi_c X, \chi_c\to \jp\;\gamma$ and $B\to\psi'X,\psi'\to\jp\;\ppc$. After combining these contributions with the two-body modes, there naturally remains a shortfall across 
a large range of $p_\psi$,
having not counted contributions from higher kaon resonances and non-resonant multi-particle final states.

\begin{figure}
\begin{minipage}[t]{7.5truecm}
\centering
\includegraphics[width=7.5truecm]{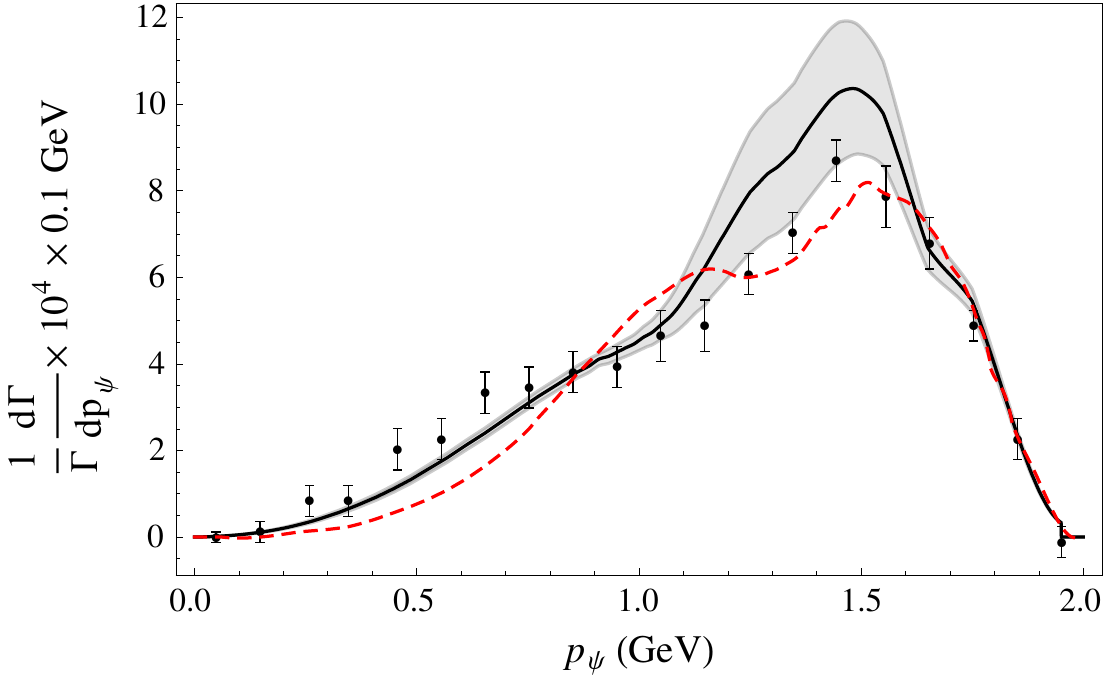}
\caption{The black-solid line represents the sum of all the contributions reported in Fig.~\ref{fig:diff2}, namely $B\to \mathcal{K}J/\psi$, $B\to\mathcal{K}\mathcal{X}$ and the octet contribution for $\Lambda_{_{\rm QCD}}=500~{\rm MeV}$ and $p_{_F}=500$~MeV, compared to the old theoretical prediction~\cite{Aubert:2002hc} (red-dashed line) computed as the sum of $B\to KJ/\psi$, $B\to K^* J/\psi$ and the color octet component with 
$\Lambda_{_{\rm QCD}}=300~{\rm MeV}$ and $p_{_F}=300$~MeV. Going from the red-dashed line to the black-solid one the $\chi^2/{\rm DOF}$ 
improves from $60/19$ to $22/19$, having included also the theoretical errors. If one choses $\Lambda_{_{\rm QCD}}=800$~MeV and $p_{_F}=300$~MeV the best fit further improves ($\chi^2/{\rm DOF}=19/19$).  Data points, black disks, are taken from BaBar~\cite{Aubert:2002hc}.}
\label{fig:new_old}
\end{minipage}
\hspace{1truecm} 
\begin{minipage}[t]{7.5truecm} 
\centering
\includegraphics[width=7.5truecm]{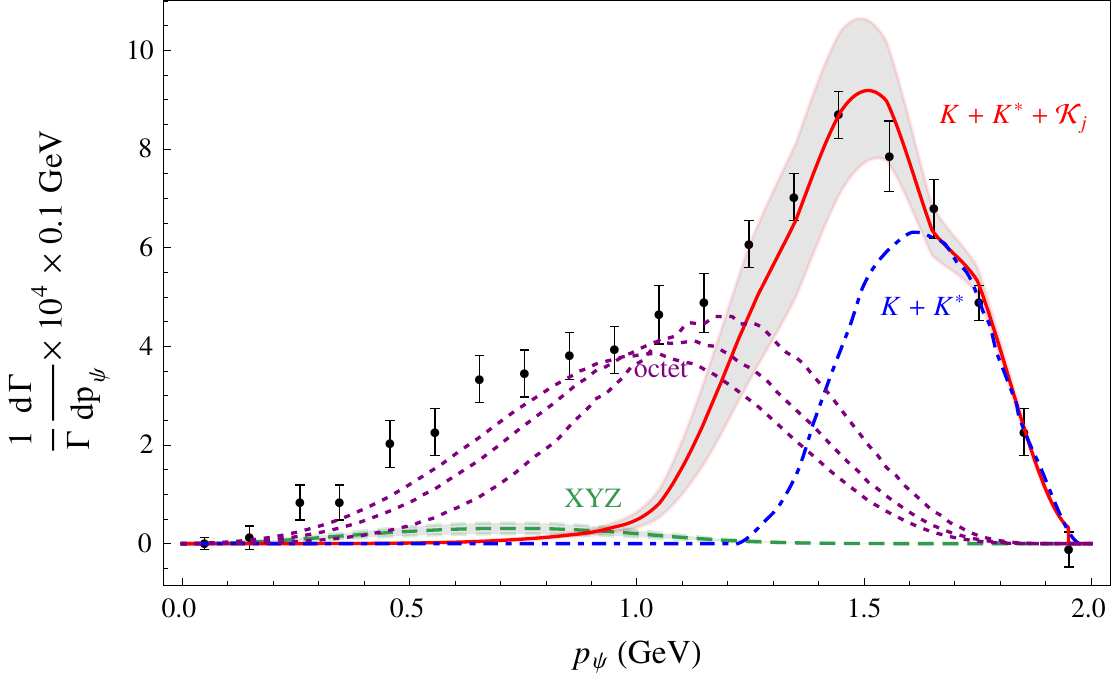}
\caption{Different contributions to the inclusive spectrum of $B\to J/\psi+{\rm All}$ decays. The red-solid line accounts for the two-body decays of the type $B\to \mathcal{K}J/\psi$;
the blue dot-dashed line shows the contributions from $B\to KJ/\psi$ and $B\to K^* J/\psi$ decays considered in~\cite{Aubert:2002hc}; the green-dashed line  represents the decays mediated by the exotic mesons $B\to\mathcal{K}\mathcal{X}\to \mathcal{K} J/\psi+{\rm light}\; {\rm hadrons}$; the purple-dotted lines come from the non-resonant multi-particle final states, {\it i.e.}, the color octet component, for three values of $\Lambda_{_{\rm QCD}}=300,500,800$~MeV (and $p_{_F}=300$~MeV) from right to left respectively, according to~\cite{Beneke:1999gq}.  Data points, black disks, are taken from BaBar~\cite{Aubert:2002hc}.}
\label{fig:diff2}
\end{minipage}
\end{figure}

Beneke \etal~\cite{Beneke:1999gq} proposed that the remaining part of  spectrum could be explained by a large $c\bar c$ color octet contribution which feeds non-resonant multi-body final states; the calculation is performed within the framework of non-relativistic-QCD (NRQCD). They confront their results with the  inclusive $\jp$ spectrum after having subtracted from it the  $B\to K\jp$ and $B\to K^*\jp$ components. The agreement found with data is rather good using reasonable values for the physical parameters involved in the computation, essentially $\Lambda_{_{\rm QCD}}$ and the Fermi momentum $p_{_F}$ of the $b$ quark inside the $B$ meson. 

Belle~\cite{Abe:2001wa} subsequently observed  the decay $B\to  K_1(1270)\jp$ with a branching fraction larger than $B\to K\jp$ and $B\to K^*\jp $. In an improved analysis of the inclusive $\jp$ spectrum, CLEO found that by summing all three of the two-body modes, a good fit to the inclusive spectra is obtained for $p_\psi>1.5$~GeV~\cite{Anderson:2002md}, which implies that the color octet contribution should be refitted.

In a higher statistics analysis by BaBar~\cite{Aubert:2002hc}, in which many feed-down modes from higher charmonia were directly measured and subtracted, the inclusive spectrum is confronted with the results of Beneke \etal, and it is noted that even after including color octet contributions, there remains an excess of events at low momenta: see Fig.~\ref{fig:new_old} 
red-dashed curve in the region $0<p_\psi<0.8$~GeV. 
This excess has provoked, since its  observation, a variety of exotic interpretations.  
In~\cite{Brodsky:1997yr} the discrepancy has been explained assuming that the $J/\psi$ recoils against a $\Lambda-\bar{p}$ strange baryonium state,
while in~\cite{Chang:2001iy,Eilam:2001kw} the existence of {\it intrinsic charm} inside the $B$-meson has been considered.
Nevertheless the most explored possibility is the production of a strange hybrid meson $K_H=s\bar{d}g$ together with the $J/\psi$ in the $B$-meson decay.
First proposed in~\cite{Eilam:2001kw}, its contribution has been later quantified in~\cite{Close:2003fz,Close:2003ae,Chua:2003fp}.
The BaBar analysis, however, does not include the large $K_1(1270)\jp$ mode, thus, for the reasons outlined above, it does not yield the full picture. 

More recently, Belle~\cite{:2010if} analyzed the large $B^+\to \jp\;K^+\ppc$ decay and performed an amplitude analysis to determine the resonant structure of the $K^+\ppc$ system, identifying several heavy kaon resonances ${\mathcal K } = K_1(1270), K_1(1400), K^*(1410), K_2^*(1430), K_2(1600), K_2(1770), K_2(1980)$.
The two-body modes $B\to {\mathcal K}J/\psi$ further constrain the contribution from non-resonant (color octet) final states, and moreover, given  
that these newly identified kaons have  large masses, they are expected to make significant contributions to the lower side of the momentum spectrum of $\jp$.  The red-solid line in  Fig.~\ref{fig:diff2}  is our picture of all the above listed exclusive two-body contributions. Not including the color octet component, the agreement with data we found extends down to $p_\psi\sim 1.2$~GeV. 

We also notice that $XYZ$ exotic resonances are found to fill the low $p_\psi$ bins as shown in Fig.~\ref{fig:diff2}: the green-dashed curve. Even if their weight in terms of branching ratio turns out to be modest with respect to what would be needed, their distribution peaks in the right region as commented in~\cite{Bigi:2005fr}.  

We thus reconsider the role of the NRQCD based model by Beneke \etal~\cite{Beneke:1999gq} with the aim of fully reconstructing the entire $d\Gamma/d{p_{\psi}}$ distribution. 
In the following we discuss our computation of the black-solid  curve in Fig.~\ref{fig:new_old} which includes a new fit  of the color octet component 
(see octet curves in Fig.~\ref{fig:diff2}) and draw our conclusions. Essentially we notice that if new states of the $XYZ$ kind will be found,
since the fit turns out to be sensitive to this component, they would aid the color octet contribution to consistently fill the gap in the data.
The observed relative small weight of $XYZ$ in terms of branching ratios, as  shown in Fig.~\ref{fig:diff2}, requires instead to force
the NRQCD calculation in the high allowed region of values for $\Lambda_{_{\rm QCD}}$ and $p_{_F}$.
 
{\bf \emph{The heavy Kaons and XYZ contributions}}.
Belle~\cite{:2010if} measures the branching ratio
\be
\mathcal{B}_{{\rm tot}}=\mathcal{B}(B^+\to J/\psi\; K^+\pi^+\pi^-)=(71.6\pm1\pm6)\times 10^{-5}
\ee
Looking at the invariant mass spectrum of $K^+\pi^+\pi^-$, they are able to isolate the resonant contributions in which $K^+\pi^+\pi^-$ 
originate from the decay of a heavy kaon $\mathcal{K}_j$ through some intermediate 
resonant state, namely $\mathcal{R}_i=K\rho/\omega, K^*\pi, K^*_0(1430)\pi, K_2^*(1430)\pi, Kf_{0,2}$, as in
\be
\mathcal{B}(B^+\to \mathcal{K}_jJ/\psi\to \mathcal{R}_iJ/\psi\to J/\psi\; K^+\pi^+\pi^-)=\mathcal{B}_{{\rm tot}} f^j_i
\ee
with $f^j_i$ the measured fractions.
Since the interference between different heavy kaons is neglected, the sum of the measured fractions is larger than one. 
The neglected interference contributions are expected to be irrelevant when $\Gamma_{\mathcal{K}_j} / m_{\mathcal{K}_j} << 1$,  
so we rescale each fraction $f^j_i$ proportionally to the width of the heavy kaon $\mathcal{K}_j$
\be
\tilde{f}^j_i=C \times\left(1-\frac{\Gamma_j}{m_j}\right)f^j_i,
\ee
where $C$ is chosen in such a way that the sum over $i$ and $j$ of $\tilde{f}^j_i$ adds up to one. Now we observe that
\be
\mathcal{B}(B^+\to \mathcal{K}_jJ/\psi\to \mathcal{R}_i J/\psi\to J/\psi\; K^+\pi^+\pi^-)=\mathcal{I}_i\times\mathcal{B}(B^+\to \mathcal{K}_jJ/\psi)\times \mathcal{B}(\mathcal{K}_j\to \mathcal{R}_i)\times \mathcal{B}(\mathcal{R}_i\to K\pi\pi),
\ee
where $\mathcal{I}_i$ are isospin factors~\footnote{$\mathcal{I}(K\rho)=1/3,\;\;\mathcal{I}(K^*\pi)=\mathcal{I}(K^*_{0,2}(1430)\pi)=4/9,\;\;\mathcal{I}(K\omega)=1,\;\;\mathcal{I}(Kf_0)=\mathcal{I}(Kf_2)=2/3$.}.  $\mathcal{B}(\mathcal{R}_i\to K\pi\pi)$ are all unity except for $\mathcal{B}(K\omega\to K\pi\pi)=0.0153$ and $\mathcal{B}(Kf_2\to K\pi\pi)=0.848$.
For some of the heavy kaons, namely $K_2(1600)$, $K_2(1770)$ and $K_2(1980)$, the values of $\mathcal{B}(\mathcal{K}_j\to \mathcal{R}_i)$ are not known experimentally and we thus extract a maximum value for them and in turn a minimum value for the relative $\mathcal{B}(B\to \mathcal{K}_jJ/\psi)$. The results obtained are summarized in Table~\ref{tab:res}.

\begin{table}[!h]
\begin{tabular}{||c|c|c|c|c||}
\hline
$ \mathcal{K}_j$ & $m_{\mathcal{K}_j}~({\rm GeV})$& $\Gamma_{\mathcal{K}_j}~({\rm GeV})$ & $\mathcal{B}(B^+\to \mathcal{K}_jJ/\psi)\times10^{5}$ \\\hline
$K_1(1270)$ & $1.270$ & $0.090$ & $144.0\pm29.3$\\
\hline
$K_1(1400)$ & $1.403$ & $0.174$ & $25.1\pm5.7$\\
\hline
$K^*(1410)$ & $1.414$ & $0.232$ & $ > 5.1\pm2.4\;\;{\rm and}\;\; <11.8\pm5.7$\\
\hline
$K^*_2(1430)$ & $1.430$ & $0.100$ & $40.2\pm 24.0$\\
\hline
$K_2(1600)$ & $1.605$ & $0.115$ & $>8.4\pm2.9$\\
\hline
$K_2(1770)$ & $1.773$ & $0.186$ & $>4.4\pm1.5$\\
\hline
$K_2(1980)$ & $1.973$ & $0.373$ & $>15.2\pm2.5$\\
\hline
\end{tabular}
\caption{Branching ratios for the decays $B\to \mathcal{K}_jJ/\psi$ extracted from Belle~\cite{:2010if}.}
\label{tab:res}
\end{table}

The branching ratios in Table~\ref{tab:res} are used to perform a Monte Carlo simulation of the decay chain, 
weighting vertices with the appropriate powers of momenta required by the decay partial wave~\footnote{Since the heaviest kaons are spin 2 states the decay occurs in {\it P}-wave and thus events are naturally pushed towards higher values of $p_\psi$ in the allowed kinematic region.}.
The sum of all contributions is shown by the red-solid curve of Fig.~\ref{fig:diff2}.

We do the same for the exotic $XYZ$ states. 
Even if almost all of these mesons decay into final states
containing a charmonium state together with light hadrons,
they cannot be easily interpreted as standard charmonia and thus are referred to as exotics~\cite{SpecNC}.
Some of them have been observed in $B$ decays produced together with the pseudoscalar kaon $K$.
The relative branching ratios are reported in Table~{\ref{tab:datixyz}}.
\begin{table}[!h]
\begin{tabular}{||c|c|c|c|c||}
\hline
${\mathcal X}_j$   & $m_{{\mathcal X}_j}~({\rm GeV})$ & $\Gamma_{{\mathcal X}_j}~({\rm GeV})$ & ${\rm Final}\;\rm{State}$ & $\mathcal{B}(B\to K{\mathcal X}_j\to K J/\psi+{\rm light}\; {\rm hadrons})\times10^{5}$\\
\hline
\multirow{2}{*}{$X(3872)$} & \multirow{2}{*}{$3.872$} & \multirow{2}{*}{$0.003$} &$J/\psi\;\rho\to J/\psi\;\pi^+\pi^-$ & $0.72\pm0.22$~\cite{Aubert:2008gu}\\ \cline{4-5}
&&&$J/\psi\;\omega$ & $0.6\pm0.3$~\cite{delAmoSanchez:2010jr} \\ \cline{4-5}
\hline
$Y(3940)$ &$3.940$ & $0.087$ &$J/\psi\;\omega$ & $3.70\pm1.14$~\cite{hfag}\\
\hline
$Y(4140)$ &$4.140$ & $0.012$ &$J/\psi\;\phi$ & $0.9\pm0.4$~\cite{Yi:2009jj}\\
\hline
$Y(4260)$ &$4.260$ & $0.095$ &$J/\psi\; f_0\to J/\psi\;\pi^+\pi^-$ & $2.00\pm0.73$~\cite{hfag}\\
\hline
\end{tabular}
\caption{Measured branching ratios for the decay $B\to K\mathcal{X}_j$. Where data is available for both neutral and charged $B$ we take the average.}
\label{tab:datixyz}
\end{table}In addition to  these decays one can suppose that $XYZ$ are produced with heavier kaons, when allowed by kinematics.
Assuming that all the exotic particles $\mathcal{X}$ are spin $1$ states (most of them do not yet have definitive quantum numbers) 
the transition matrix elements are given by $\langle \mathcal{X}(\epsilon,p)\mathcal{K}(q)|B(P)\rangle=g\;\epsilon\cdot q$
for a spin $0$ kaon and by $\langle \mathcal{X}(\epsilon,p)\mathcal{K}(\eta,q)|B(P)\rangle=g^\prime\;\epsilon\cdot \eta$ for a spin $1$ kaon\footnote{We cannot consider here the decay $B\to\mathcal{K}\mathcal{X}$ with a spin 2 kaon, since we do not have any experimental data to compare with.}.
From a dimensional analysis $[g]=M^0$ and $[g^\prime]=M$. Since we only have data on the branching ratios 
$\mathcal{B}(B\to K\mathcal{X})$ we need some hypothesis on the relationship between $g$ and $g^\prime$.
We suppose that $g'=\Lambda\;g$, with $\Lambda$ some mass scale. 
Using $\mathcal{B}(B\to KX(3872))$ reported in Table~\ref{tab:datixyz} and the upper limit~\cite{:2008te}
\be
\mathcal{B}(B\to K^*X(3872))\times \mathcal{B}(X(3872)\to J/\psi\;\pi^+\pi^-)<0.34\times 10^{-5}
\ee
we obtain $\Lambda\gtrsim 550$~MeV. We thus decide to extract $g$ from the known branching ratios reported in Table~\ref{tab:datixyz}
and fix $g^\prime=m_{_{\mathcal{K}(J=1)}}g$, where $m{_\mathcal{K}(J=1)}$ is the mass of a generic spin $1$ kaon.
The green-dashed line in Fig.~\ref{fig:diff2} shows the sum of all $XYZ$ contributions along with all kaons allowed by kinematics.

{\bf \emph{The color octet component}}.
We reanalyzed the Beneke \etal~\cite{Beneke:1999gq} calculation to refit the color octet curve contribution to the $J/\psi$ spectrum.
Their computation, based on the NRQCD factorization approach, assumes that the production of the $c\bar{c}$ pair in a color octet state, 
which has approximately the same probability as in color singlet, has a large kinematic effect on the momentum spectrum, 
due to the energy distribution of the soft gluons emitted by the $(c\bar c)_{\bf 8}$ fragmentation into $J/\psi$. 

The $(c\bar c)_{\bf 8}$ pair can be produced, at order $v^4$ in the non relativistic expansion, in three different $L-S$ configurations: $1^1\!S_0$, $1^3\!P_J$ and $1^3\!S_1$.
One can associate with each of these configurations
a non-perturbative shape function for the energy and an invariant mass distribution of the radiated system 
(with a characteristic energy scale of $m_c v^2\approx\Lambda_{_{\rm QCD}}$). 
The normalizations of the shape functions are related to the conventional NRQCD matrix elements $\langle \mathcal{O}^{J/\psi}_n\rangle$, which describe the hadronization into a $J/\psi$ from a $c\bar c$ pair in an angular momentum and color state $n$. 
The $\langle \mathcal{O}^{J/\psi}_n\rangle$ can be determined by fits to $J/\psi$ production in a variety of processes.
Nevertheless, enforcing these normalization conditions underestimates data, because the phenomenological values of the matrix elements are computed from integrated quantities in leading order calculations, while the shape functions contain higher order corrections in the velocity expansion.
To overcome this difficulty one chooses to fix the absolute normalization by adjusting the sum of all contributions to data.
The Fermi motion of the $b$ quark inside the $B$ meson is taken into account using the ACCMM model~\cite{Altarelli:1982kh}. The tunable parameters are essentially $\Lambda_{_{\rm QCD}}$ and the Fermi momentum $p_{_F}$.

In this approach the color octet channel leads to non-resonant multi-body final states since  the probability that the emitted soft gluons reassemble with the spectators (light quarks and gluons inside the $B$) to form a single hadron is assumed to be very small (factorization hypothesis).  
The two-body modes are due to the color singlet channel and must be considered separately.

In~\cite{Beneke:1999gq} the comparison is made with data where the two-body modes with $K$ and $K^*$ are subtracted. The agreement with the experimental spectrum,
having fixed  $\Lambda_{_{\rm QCD}}=300$~MeV and $p_{_F}=300$~MeV, is shown in Fig.~\ref{fig:new_old}, red-dashed curve.

An observation is in order: possible interference effects between the octet final states, which at least contain two pions in addition to  $\mathcal{K}J/\psi$, 
and the multi-body final states originated from the exclusive modes with heavier kaons decaying to  ${\cal K} + $~pions and $XYZ$ resonances 
can be safely neglected as $\Gamma_{{\cal K}_j} / m_{{\cal K}_j} < 0.2$ and $\Gamma_{{\cal X}_j} / m_{{\cal X}_j} << 1$. 
This would be a stronger assumption for hypothetical new broad resonances. 

{\bf \emph{Results and outlook}}.
The black-solid curve in Fig.~\ref{fig:new_old} is obtained as a sum of the standard two-body contributions from kaons (red-solid curve 
in Fig.~\ref{fig:diff2}) with the $XYZ$  contributions (green-dashed) plus one of the color octet components (purple-dotted). 
The latter are three curves obtained with $\Lambda_{_{\rm QCD}}=300,500,800$~MeV (and $p_{_F}=300$~MeV) from right 
to left respectively. 
To best fit data in Fig.~\ref{fig:new_old} we choose the octet component with $\Lambda_{_{\rm QCD}}=500$~MeV 
and also need to push $p_{_F}$ up to $p_{_F}=500$~MeV. 
The values chosen for $\Lambda_{_{\rm QCD}}$ and $p_{_F}$  are critically on the high sides of the allowed ranges
which are supposed to be $\Lambda_{_{\rm QCD}}\in [200,450]$~MeV~\cite{Bodwin:1994jh} and $p_{_F}\in [300,450]$~MeV. 
Yet the black-solid curve in Fig.~\ref{fig:new_old} represents a considerable improvement 
with respect to the old one (red-dashed). Relying on the validity of the NRQCD approach, our results seem to indicate that 
the addition of new resonances of the $XYZ$ kind feeding the low $p_\psi$ region would effectively improve the agreement with data.  
Indeed if the total branching ratio due to the $XYZ$ turns out to be three times the currently observed value, a good description
of data, namely $\chi^2/{\rm DOF}=28/19$, would be obtained including a color octet component with $\Lambda_{_{\rm QCD}}=300$~MeV
and $p_{_F}=500$~MeV.
In this respect our results could be suggestive of the existence of a number of not yet discovered exotic mesons:
new outcomes may arrive from Belle and LHCb.

\begin{acknowledgments}
We thank R.~Faccini for interesting discussions and useful information on the experimental data.
\end{acknowledgments}

\bibliography{JPsiMomentum}

\begin{thebibliography}{22}
\expandafter\ifx\csname natexlab\endcsname\relax\def\natexlab#1{#1}\fi
\expandafter\ifx\csname bibnamefont\endcsname\relax
  \def\bibnamefont#1{#1}\fi
\expandafter\ifx\csname bibfnamefont\endcsname\relax
  \def\bibfnamefont#1{#1}\fi
\expandafter\ifx\csname citenamefont\endcsname\relax
  \def\citenamefont#1{#1}\fi
\expandafter\ifx\csname url\endcsname\relax
  \def\url#1{\texttt{#1}}\fi
\expandafter\ifx\csname urlprefix\endcsname\relax\def\urlprefix{URL }\fi
\providecommand{\bibinfo}[2]{#2}
\providecommand{\eprint}[2][]{\url{#2}}

\bibitem[{\citenamefont{Balest et~al.}(1995)}]{Balest:1994jf}
\bibinfo{author}{\bibfnamefont{R.}~\bibnamefont{Balest}} \bibnamefont{et~al.}
  (\bibinfo{collaboration}{CLEO}), \bibinfo{journal}{Phys. Rev.}
  \textbf{\bibinfo{volume}{D52}}, \bibinfo{pages}{2661} (\bibinfo{year}{1995}).

\bibitem[{\citenamefont{Alam et~al.}(1994)}]{Alam:1994bi}
\bibinfo{author}{\bibfnamefont{M.~S.} \bibnamefont{Alam}} \bibnamefont{et~al.}
  (\bibinfo{collaboration}{CLEO}), \bibinfo{journal}{Phys. Rev.}
  \textbf{\bibinfo{volume}{D50}}, \bibinfo{pages}{43} (\bibinfo{year}{1994}),
  \eprint{hep-ph/9403295}.

\bibitem[{\citenamefont{Aubert et~al.}(2003)}]{Aubert:2002hc}
\bibinfo{author}{\bibfnamefont{B.}~\bibnamefont{Aubert}} \bibnamefont{et~al.}
  (\bibinfo{collaboration}{BABAR}), \bibinfo{journal}{Phys. Rev.}
  \textbf{\bibinfo{volume}{D67}}, \bibinfo{pages}{032002}
  (\bibinfo{year}{2003}), \eprint{hep-ex/0207097}.

\bibitem[{\citenamefont{Beneke et~al.}(2000)\citenamefont{Beneke, Schuler, and
  Wolf}}]{Beneke:1999gq}
\bibinfo{author}{\bibfnamefont{M.}~\bibnamefont{Beneke}},
  \bibinfo{author}{\bibfnamefont{G.~A.} \bibnamefont{Schuler}},
  \bibnamefont{and} \bibinfo{author}{\bibfnamefont{S.}~\bibnamefont{Wolf}},
  \bibinfo{journal}{Phys. Rev.} \textbf{\bibinfo{volume}{D62}},
  \bibinfo{pages}{034004} (\bibinfo{year}{2000}), \eprint{hep-ph/0001062}.

\bibitem[{\citenamefont{Abe et~al.}(2001)}]{Abe:2001wa}
\bibinfo{author}{\bibfnamefont{K.}~\bibnamefont{Abe}} \bibnamefont{et~al.}
  (\bibinfo{collaboration}{Belle}), \bibinfo{journal}{Phys. Rev. Lett.}
  \textbf{\bibinfo{volume}{87}}, \bibinfo{pages}{161601}
  (\bibinfo{year}{2001}), \eprint{hep-ex/0105014}.

\bibitem[{\citenamefont{Anderson et~al.}(2002)}]{Anderson:2002md}
\bibinfo{author}{\bibfnamefont{S.}~\bibnamefont{Anderson}} \bibnamefont{et~al.}
  (\bibinfo{collaboration}{CLEO}), \bibinfo{journal}{Phys. Rev. Lett.}
  \textbf{\bibinfo{volume}{89}}, \bibinfo{pages}{282001}
  (\bibinfo{year}{2002}).

\bibitem[{\citenamefont{Brodsky and Navarra}(1997)}]{Brodsky:1997yr}
\bibinfo{author}{\bibfnamefont{S.~J.} \bibnamefont{Brodsky}} \bibnamefont{and}
  \bibinfo{author}{\bibfnamefont{F.~S.} \bibnamefont{Navarra}},
  \bibinfo{journal}{Phys. Lett.} \textbf{\bibinfo{volume}{B411}},
  \bibinfo{pages}{152} (\bibinfo{year}{1997}), \eprint{hep-ph/9704348}.

\bibitem[{\citenamefont{Chang and Hou}(2001)}]{Chang:2001iy}
\bibinfo{author}{\bibfnamefont{C.-H.~V.} \bibnamefont{Chang}} \bibnamefont{and}
  \bibinfo{author}{\bibfnamefont{W.-S.} \bibnamefont{Hou}},
  \bibinfo{journal}{Phys. Rev.} \textbf{\bibinfo{volume}{D64}},
  \bibinfo{pages}{071501} (\bibinfo{year}{2001}), \eprint{hep-ph/0101162}.

\bibitem[{\citenamefont{Eilam et~al.}(2002)\citenamefont{Eilam, Ladisa, and
  Yang}}]{Eilam:2001kw}
\bibinfo{author}{\bibfnamefont{G.}~\bibnamefont{Eilam}},
  \bibinfo{author}{\bibfnamefont{M.}~\bibnamefont{Ladisa}}, \bibnamefont{and}
  \bibinfo{author}{\bibfnamefont{Y.-D.} \bibnamefont{Yang}},
  \bibinfo{journal}{Phys. Rev.} \textbf{\bibinfo{volume}{D65}},
  \bibinfo{pages}{037504} (\bibinfo{year}{2002}), \eprint{hep-ph/0107043}.

\bibitem[{\citenamefont{Close and Dudek}(2003)}]{Close:2003fz}
\bibinfo{author}{\bibfnamefont{F.~E.} \bibnamefont{Close}} \bibnamefont{and}
  \bibinfo{author}{\bibfnamefont{J.~J.} \bibnamefont{Dudek}},
  \bibinfo{journal}{Phys. Rev. Lett.} \textbf{\bibinfo{volume}{91}},
  \bibinfo{pages}{142001} (\bibinfo{year}{2003}), \eprint{hep-ph/0304243}.

\bibitem[{\citenamefont{Close and Dudek}(2004)}]{Close:2003ae}
\bibinfo{author}{\bibfnamefont{F.~E.} \bibnamefont{Close}} \bibnamefont{and}
  \bibinfo{author}{\bibfnamefont{J.~J.} \bibnamefont{Dudek}},
  \bibinfo{journal}{Phys. Rev.} \textbf{\bibinfo{volume}{D69}},
  \bibinfo{pages}{034010} (\bibinfo{year}{2004}), \eprint{hep-ph/0308098}.

\bibitem[{\citenamefont{Chua et~al.}(2003)\citenamefont{Chua, Hou, and
  Wong}}]{Chua:2003fp}
\bibinfo{author}{\bibfnamefont{C.-K.} \bibnamefont{Chua}},
  \bibinfo{author}{\bibfnamefont{W.-S.} \bibnamefont{Hou}}, \bibnamefont{and}
  \bibinfo{author}{\bibfnamefont{G.-G.} \bibnamefont{Wong}},
  \bibinfo{journal}{Phys. Rev.} \textbf{\bibinfo{volume}{D68}},
  \bibinfo{pages}{054012} (\bibinfo{year}{2003}), \eprint{hep-ph/0305180}.

\bibitem[{\citenamefont{Guler et~al.}(2011)}]{:2010if}
\bibinfo{author}{\bibfnamefont{H.}~\bibnamefont{Guler}} \bibnamefont{et~al.}
  (\bibinfo{collaboration}{Belle}), \bibinfo{journal}{Phys. Rev.}
  \textbf{\bibinfo{volume}{D83}}, \bibinfo{pages}{032005}
  (\bibinfo{year}{2011}), \eprint{1009.5256}.

\bibitem[{\citenamefont{Bigi et~al.}(2005)\citenamefont{Bigi, Maiani,
  Piccinini, Polosa, and Riquer}}]{Bigi:2005fr}
\bibinfo{author}{\bibfnamefont{I.}~\bibnamefont{Bigi}},
  \bibinfo{author}{\bibfnamefont{L.}~\bibnamefont{Maiani}},
  \bibinfo{author}{\bibfnamefont{F.}~\bibnamefont{Piccinini}},
  \bibinfo{author}{\bibfnamefont{A.~D.} \bibnamefont{Polosa}},
  \bibnamefont{and} \bibinfo{author}{\bibfnamefont{V.}~\bibnamefont{Riquer}},
  \bibinfo{journal}{Phys. Rev.} \textbf{\bibinfo{volume}{D72}},
  \bibinfo{pages}{114016} (\bibinfo{year}{2005}), \eprint{hep-ph/0510307}.

\bibitem[{\citenamefont{Drenska et~al.}(2010)\citenamefont{Drenska, Faccini,
  Piccinini, Polosa, Renga et~al.}}]{SpecNC}
\bibinfo{author}{\bibfnamefont{N.}~\bibnamefont{Drenska}},
  \bibinfo{author}{\bibfnamefont{R.}~\bibnamefont{Faccini}},
  \bibinfo{author}{\bibfnamefont{F.}~\bibnamefont{Piccinini}},
  \bibinfo{author}{\bibfnamefont{A.}~\bibnamefont{Polosa}},
  \bibinfo{author}{\bibfnamefont{F.}~\bibnamefont{Renga}},
  \bibnamefont{et~al.}, \bibinfo{journal}{Riv.Nuovo Cim.}
  \textbf{\bibinfo{volume}{033}}, \bibinfo{pages}{633} (\bibinfo{year}{2010}),
  \eprint{1006.2741}.

\bibitem[{\citenamefont{Aubert et~al.}(2008)}]{Aubert:2008gu}
\bibinfo{author}{\bibfnamefont{B.}~\bibnamefont{Aubert}} \bibnamefont{et~al.}
  (\bibinfo{collaboration}{BABAR}), \bibinfo{journal}{Phys. Rev.}
  \textbf{\bibinfo{volume}{D77}}, \bibinfo{pages}{111101}
  (\bibinfo{year}{2008}), \eprint{0803.2838}.

\bibitem[{\citenamefont{del Amo~Sanchez et~al.}(2010)}]{delAmoSanchez:2010jr}
\bibinfo{author}{\bibfnamefont{P.}~\bibnamefont{del Amo~Sanchez}}
  \bibnamefont{et~al.} (\bibinfo{collaboration}{BABAR}),
  \bibinfo{journal}{Phys. Rev.} \textbf{\bibinfo{volume}{D82}},
  \bibinfo{pages}{011101} (\bibinfo{year}{2010}), \eprint{1005.5190}.

\bibitem[{hfa()}]{hfag}
\bibinfo{howpublished}{\url{http://hfag.phys.ntu.edu.tw/b2charm/index.html}}.

\bibitem[{\citenamefont{Yi and collaboration}(2009)}]{Yi:2009jj}
\bibinfo{author}{\bibfnamefont{K.}~\bibnamefont{Yi}} \bibnamefont{and}
  \bibinfo{author}{\bibfnamefont{f.~t.~C.} \bibnamefont{collaboration}},
  \bibinfo{journal}{PoS EPS-HEP} \textbf{\bibinfo{volume}{2009}},
  \bibinfo{pages}{2009:085,2009} (\bibinfo{year}{2009}), \eprint{0910.3163}.

\bibitem[{\citenamefont{Adachi et~al.}(2008)}]{:2008te}
\bibinfo{author}{\bibfnamefont{I.}~\bibnamefont{Adachi}} \bibnamefont{et~al.}
  (\bibinfo{collaboration}{Belle}) (\bibinfo{year}{2008}), \eprint{0809.1224}.

\bibitem[{\citenamefont{Altarelli et~al.}(1982)\citenamefont{Altarelli,
  Cabibbo, Corbo, Maiani, and Martinelli}}]{Altarelli:1982kh}
\bibinfo{author}{\bibfnamefont{G.}~\bibnamefont{Altarelli}},
  \bibinfo{author}{\bibfnamefont{N.}~\bibnamefont{Cabibbo}},
  \bibinfo{author}{\bibfnamefont{G.}~\bibnamefont{Corbo}},
  \bibinfo{author}{\bibfnamefont{L.}~\bibnamefont{Maiani}}, \bibnamefont{and}
  \bibinfo{author}{\bibfnamefont{G.}~\bibnamefont{Martinelli}},
  \bibinfo{journal}{Nucl.Phys.} \textbf{\bibinfo{volume}{B208}},
  \bibinfo{pages}{365} (\bibinfo{year}{1982}).

\bibitem[{\citenamefont{Bodwin et~al.}(1995)\citenamefont{Bodwin, Braaten, and
  Lepage}}]{Bodwin:1994jh}
\bibinfo{author}{\bibfnamefont{G.~T.} \bibnamefont{Bodwin}},
  \bibinfo{author}{\bibfnamefont{E.}~\bibnamefont{Braaten}}, \bibnamefont{and}
  \bibinfo{author}{\bibfnamefont{G.}~\bibnamefont{Lepage}},
  \bibinfo{journal}{Phys.Rev.} \textbf{\bibinfo{volume}{D51}},
  \bibinfo{pages}{1125} (\bibinfo{year}{1995}), \eprint{hep-ph/9407339}.

\end{thebibliography}

\bigskip 

\end{document}